\documentclass[aps,prb,superscriptaddress,runinaddress,floatfix]{revtex4}

\usepackage{eurosym,amsfonts,amssymb,amsmath,times}
\usepackage{bm}
\usepackage{graphicx}
\usepackage{dcolumn}
\usepackage{hyperref}
\hypersetup{colorlinks=true,linkcolor=blue,citecolor=blue,urlcolor=blue}
\usepackage[letterpaper,left=0.7in,right=0.7in,top=0.75in,bottom=0.55in]{geometry}

\begin{document}

\title{Kinetic effects observed in dynamic FORCs of magnetic wires.
Experiment and theoretical description.}
\author{Dorin Cimpoesu}
\email{cdorin@uaic.ro}
\author{Ioan Dumitru}
\author{Alexandru Stancu}
\affiliation{Department of Physics, Alexandru Ioan Cuza University of Iasi, Iasi 700506,
Romania}

\begin{abstract}
This study is focused on the possibility to extend the use of the
first-order reversal curve (FORC) diagram method to rate-dependent
hysteresis. The FORCs for an amorphous magnetic wire was measured with an
inductometric experimental setup in which the field-rate was maintained
constant. The FORC experiment was performed for four different field-rates.
As it is known, to obtain quantitative information on the magnetization
process during the FORC process we need a model able to simulate as close as
possible the experimental FORC diagrams. In this case, we have developed and
implemented a model based on the hypothesis that the magnetization processes
in this kind of materials are mainly due to the movement of a domain wall
between the central domains of the wire. The differential equation of the
domain wall movement is able to give a remarkably accurate description of
the experimental FORC diagrams. The experimental FORCs, the FORC
susceptibility diagram and the classical FORC diagram show however a number
of details that the model is not able to describe. In each such case one
discuss the possible physical cause of the observed behavior. As the
magnetic wires are analyzed in many laboratories around the world for a wide
variety of applications (essentially involving the control of the domain
wall movement) we consider that our study offers to these researchers a
valuable new tool.
\end{abstract}

\maketitle

\section{Introduction}

One of the most successful new characterization technique, introduced in
magnetism a few decades ago, known as the FORC diagram method, is gradually
replacing the measurement of the major hysteresis loop that was predominant
for a long time in the scientific literature. Originated in theoretical
studies of ferromagnetic hysteresis based on the Classical Preisach Model
(CPM) \cite{Preisach 1935} the idea of using first-order reversal curves
(FORC) as a non-parametrical identification technique for this model was
first published by Mayergoyz in 1985.\cite{Mayergoyz JAP 1985} As stated in
this article, the most important point in using FORCs in the process of
identification is that all the magnetic states of the systems are well
defined in this type of experiment. However, originally the method was only
recommended for the identification of the parameters of the CPM, that is, if
and only if the hysteretic process has two properties: congruency of the
minor loops measured within the same field limits and the wiping-out of the
system memory after a closed minor loop (these systems are named CPM
systems). Unfortunately most magnetic systems don't obey to these two
conditions and consequently it didn't made too much sense to
experimentalists to continue with the FORC measurement after they observed
large disagreements from the congruency and/or wiping-out properties. This
Gordian knot has been cut by a group \cite{Pike JAP 1999} from Davis
University when they introduced a really efficient numerical tool to
calculate a distribution from a set of FORCs covering the major hysteresis
loop for a variety of samples and obtained in a systematic manner what they
called FORC distributions for these samples. The fundamental new idea at
that moment was that one can use the FORC distributions and the FORC
diagrams, which are contour plots of these distributions, as tools of
magnetic characterization for the samples. They have observed that the
magnetic samples for a given category of magnetic material or system are
similar but not identical and they introduced the idea of \textquotedblleft
fingerprinting\textquotedblright\ the material using FORC diagrams, idea
which was really successful. As the authors claimed that the FORC method is
not an identification technique for the Preisach model, the problem faced by
the users of the method was how to interpret quantitatively these
\textquotedblleft fingerprints\textquotedblright , if at all possible.
Systematic studies however have shown that it is counterproductive to deny
the profound link between the FORC diagram method and the Preisach model 
\cite{Stancu JAP 2003} and that modified Preisach-type models like the
moving Preisach model \cite{Della Torre 1965} could produce FORC diagrams
very similar to a variety of experimental ones. This fact indicated that the
FORC diagram method could become a quantitative method in some conditions.
The main condition to transform the FORC method from a virtually qualitative
one to a sound quantitative method is to find a model able to reproduce,
using some fitting physical parameters, the experimental diagrams. Depending
on the quality of this model, of its ability to describe not only one case
but also systematic series of samples, the physical interpretation of the
FORC diagrams becomes more and more trustworthy. As shown in Ref. %
\onlinecite{Stancu JAP 2003}, many times the model can be a modified version
of the Classical Preisach Model.

Nevertheless, many FORC diagram users are not employing properly this tandem
experimental FORC/model but still are discussing their experimental results
in quantitative terms. The most usual interpretation of the FORC
distribution is as a vaguely distorted Preisach distribution. In this way,
experimental FORC distributions are presented as estimates of the real
distribution of the magnetic elements in the sample as a function of
coercive and interaction fields. Of course, this is not exactly true and can
be misleading in many cases. Many recent studies have shown what are the
limits of this type of interpretation, especially if one expects to find
physical elements related in a biunivocal way with a certain region on the
FORC distribution.\cite{Dobrota JAP 2013, Dobrota PB 2015, Nica PB 2015}
Similar discussions are specific not only to experimental magnetism and
especially ferromagnetism, but also to the other areas of science in which
FORC diagram technique is used to characterize complex hysteretic systems,
like in ferroelectricity \cite{Stancu APL 2003}, temperature, light-induced
and pressure hysteresis in spin-transition materials \cite{Tanasa PRB 2005,
Enachescu PRB 2005, Rotaru PRB 2011, Kou AM 2011}, etc.

An important point that has to be discussed in details when quantitative
interpretation of the FORC diagram is attempted is that in most experimental
studies performed on magnetic samples showing hysteresis one assumes that
the results are not essentially modified if the experimental time is changed
within a certain margin. We also mention that most of the Preisach-type
models are also time independent. Fundamentally, this means that the output
value, which is the total magnetic moment of the studied sample, is not
changing if the field is applied a shorter or longer duration. In fact, in
experiments there are two distinct types of methods used for the measurement
of the magnetic moment versus applied magnetic field curves. In one category
of experiments the field is constant during the measurement of the magnetic
moment, as in experiments with vibrating sample magnetometers (VSM). Another
category, which is based on the electromagnetic induction phenomenon,
implies a continuous variation of the applied magnetic field which produces
a variation of the magnetic moment that is detected with appropriate systems
of detection coils. If in the first experiment one can define a clear
duration for the applied field in each experimental point, in the second one
it is better to discuss the result as being influenced by the instantaneous
value of the applied field and the field rate. Nevertheless, in both types
of experiments in many concrete cases one can observe significant changes in
the values of the measured magnetic moment as a function of experiment time
(in the magnetometric measurements) or as a function of field rate (in the
inductometric measurements). Both these phenomena could be included in the
category of kinetic effects which may affect quite dramatically the results
of measurements made on various magnetic samples. Of course these effects
are related with the magnetic relaxation of the fundamental elements in the
studied sample (particles, magnetic domains). It is also known for a long
time that the distribution of the typical relaxation times within the sample
is vital in the actual behavior of the magnetic sample as a function of the
applied field and as a function of time.\cite{Street Wooley, Neel}

To investigate the kinetic effects on magnetic samples we have selected as
samples magnetic wires. Isolated magnetic wires or systems of interacting
wires are at this moment perhaps the most promising magnetic materials for a
wide variety of applications. In a recently published monography \cite%
{Vazquez 2015} dedicated to the magnetic nano- and microwires one can find
an illustration not only of the diversity of design and synthesis methods,
but also of the wide range of applications for these materials. The first
magnetic characterization of nanowire systems with the static FORC technique
was published in 2004 \cite{Spinu IEEE 2004} and since then many groups have
concentrated their attention on similar types of samples. \cite{Clime JAP
2007, Beron book 2010, Rotaru PRB 2011, Proenca Nanotechnology 2013,
Almasi-Kashi PB 2014} In most of these articles the preferred measurement
technique was the VSM and the dependence of the results on the waiting time
in each experimental point was negligible. However, one of the most exciting
applications of the magnetic wires is in microwave technology (see Chapter
17 in Ref. \onlinecite{Vazquez
2015}). The ac excitation field, especially as the frequency is increased,
is fundamentally different from the quasistatic experiment. One certainly
expects that the FORC diagrams to be strongly dependent on the frequency of
the ac field and in this case we should unquestionably need a model to
explain this behavior. In order to make a clear distinction between the
quasistatic FORCs and the one in which the field rate is decisive in shaping
the FORC distribution, we shall name the second one dynamic FORC (dFORC)
diagram.

In this article we analyze the influence of the kinetic effects on dynamic
FORC diagrams measured in an inductometric experimental setup, designed for
magnetic wires. In the section dedicated to the experimental setup we
present all the details concerning the control of the field rate during the
FORC-type experiment. After we show the experimental results obtained for a
sample measured for several values for the field rate, we have developed a
simple model able to reproduce with a remarkable accuracy not only the
typical features for one field rate but also how these features are changing
when the field rate is changing. In this way one can provide with a high
degree of certitude what are the connections between the experimental FORCs
and physical phenomena related to the movement of the domain walls when an
ac-type of magnetic field is applied to a magnetic wire.

\section{Experimental setup}

\begin{figure}[tbp]
\includegraphics[width=75mm,keepaspectratio=true]{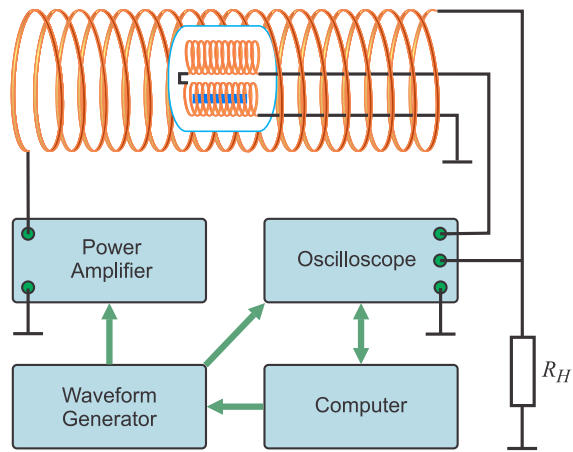}
\caption{Block diagram of the induction magnetometer}
\label{Fig_1}
\end{figure}

\begin{figure*}[tbp]
\includegraphics[width=180mm,keepaspectratio=true]{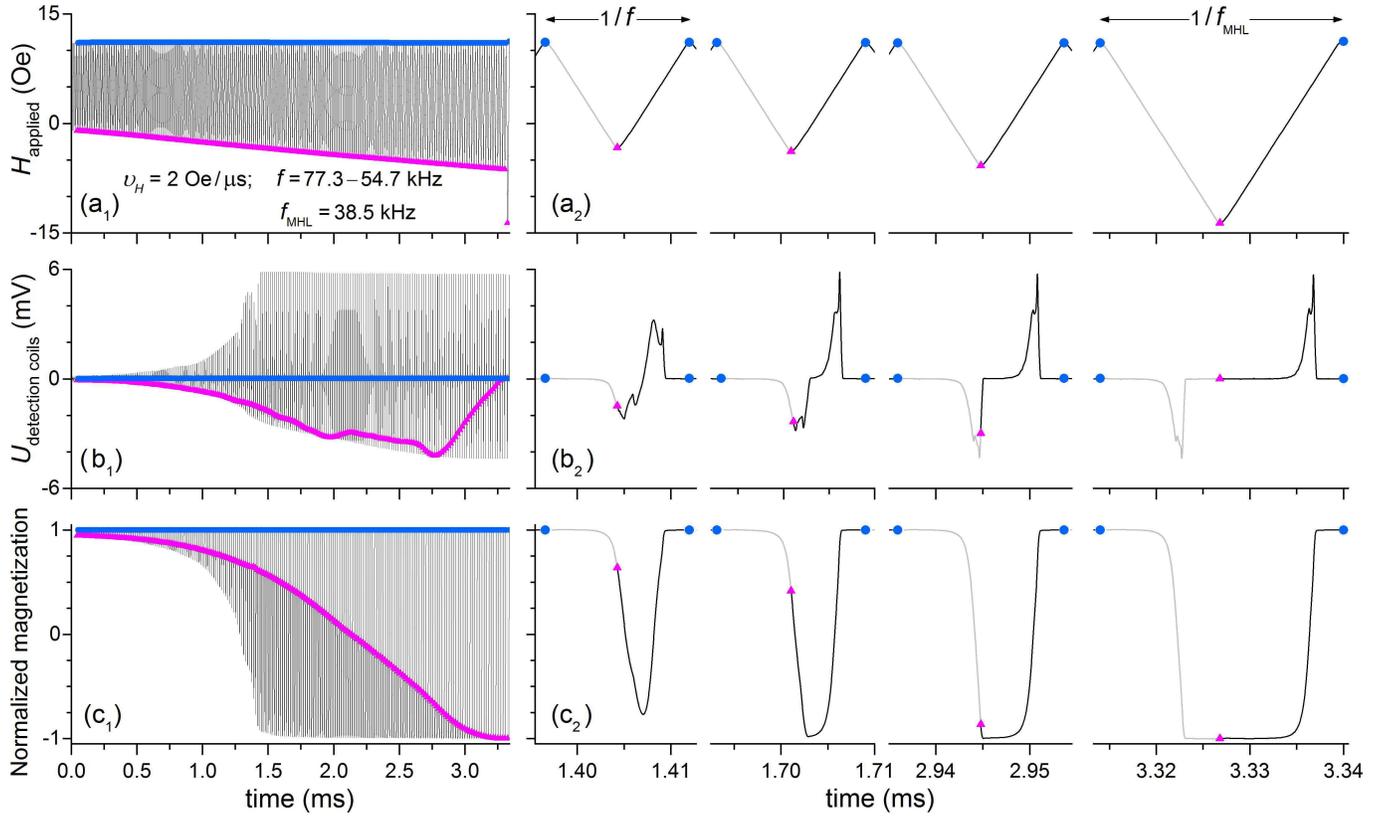}
\caption{Time variation of the applied magnetic field [(a$_{1}$) and (a$_{2}$%
)] and of the induced signal across the detection coils [(b$_{1}$) and (b$%
_{2}$)] for a value of the field sweep rate of $\protect\upsilon _{H}=2\,%
\mathrm{Oe/\protect\mu s}$ used to obtain the dynamic FORCs and FORC diagram
from Fig. \protect\ref{Fig_3}(d$_{1}$). The last FORC is only used to obtain
the dynamic major hysteresis loop (dMHL). The changing magnetic moment of
the sample is proportional with the time integration of the induced signal.
The triangle symbols mark the time moments when the applied field starts to
increase on each FORC, i.e., when it is equal to the reversal field on each
FORC, while the circles mark the time moments when the field attains its
maximum value and then begins to decrease. Portions corresponding to
decreasing fields are marked in gray. Moir\'{e}\ effect from (a$_{1}$-c$_{1}$%
) and is only an undesired visual artifact due to the proximity of the
adjacent curves.}
\label{Fig_2}
\end{figure*}

A FORC is obtained by saturating the sample under study in a positive
magnetic field $H_{\mathrm{sat}}$, decreasing the field to the reversal
field $H_{\mathrm{reversal}}$, and then sweeping the field back to $H_{%
\mathrm{sat}}$. The FORC is defined as the magnetization curve that results
when the applied field is increased from $H_{\mathrm{reversal}}$ to $H_{%
\mathrm{sat}}$, and it is a function of the applied field $H_{\mathrm{applied%
}}$ and the reversal field $H_{\mathrm{reversal}}$. This process is repeated
for many values of $H_{\mathrm{reversal}}$, so that the reversal points
cover the descending branch of the major hysteresis loop (MHL), while the
corresponding FORCs cover the hysteretic surface of the MHL. The FORC
diagram is defined as the mixed second derivative of the set of FORCs with
respect to $H_{\mathrm{applied}}$ and $H_{\mathrm{reversal}}$, taken with
negative sign.\cite{Pike JAP 1999} Before each FORC measurement, saturation
needs to be achieved in order to erase completely the previous magnetic
history.

In order to measure the dynamic FORCs, an ac magnetometer based on the
induction coil principle \cite{book induction magnetometer, Beron RSI 11}
was constructed. Two identical pickup/detection coils differentially
connected (i.e., in series opposition so that the applied magnetic field
produces no net signal/voltage) measure the sample's magnetic response by
detecting the induced voltage across the detection coils due to the changing
magnetic moment of the sample which is placed within one of the coils (see
Fig. \ref{Fig_1}). The detection coils should be as far apart to reduce both
the mutual inductive coupling between them and the coupling between the
empty detection coil and the magnetic sample. The coils geometry can be
optimized for various sample shapes. Detection coils assembly is centered in
a solenoid which provides the nearly uniform external excitation field [see
Figs. \ref{Fig_2}(a$_{1}$) and (a$_{2}$) for the time variation of the
applied field] through a function/arbitrary waveform generator (Picotest
G5100A) and a high speed, broad band (dc to $1\,\mathrm{MHz}$) and high slew
rate ($600\,\mathrm{V/\mu s}$ ) bipolar power amplifier (HSA 4011/HSA 4014
NF Corporation). The diameter of the solenoid should be few times bigger
than that of the detection coils in order to have a uniform excitation field
as well as to diminish the influence of the sample on the solenoid (i.e.,
the solenoid's inductance should not depend on the magnetic sample). A high
power and low inductance (over a wide frequency range) resistor $R_{H}$\
connected in series with the solenoid allows to measure the electric current
that flows through the solenoid and generates the external magnetic field
(we note that the inductive reactance of the usual high power resistors can
not be neglected beside its resistance even for low frequencies). The
capability of supplying high output voltage and high power of the amplifier
allow us to obtain up to $800\,\mathrm{Oe}$. For reliable measurements made
on soft magnetic materials, a Helmholtz coil assembly could be used to
compensate the Earth's magnetic field. Time variation of the applied field
(which is proportional to the voltage across the low inductance resistor $%
R_{H}$) and of the induced signal in the detection coils were digitized and
acquired through a high resolution DPO7254 Tektronix digital oscilloscope ($%
2.5\,\mathrm{GHz}$ bandwidth and $10\,\mathrm{GS}\left/ \mathrm{s}\right. $
synchronous real time sampling rate on all four input channels) synchronized
with the waveform generator, and then sent to a computer for software signal
processing. The high input impedance of the oscilloscope determines a very
low (practically zero) current in the detection coils. Average acquisition
mode was used to obtain the average value for each record point over many
acquisitions, in order to reduce random noise. To further reduce the noise
we have used the advantage of digital filtering which improves
signal-to-noise performance over analog filters, which have to sacrifice
accuracy in order to perform over a wide frequency band. No external
amplifier was used for the induced voltage, only the internal amplifier of
the oscilloscope. The high speed acquisition allows fast and detailed
dynamic FORC diagrams measurement in a wide range of the applied field sweep
rate values. All the instrumentation are interconnected with a computer via
GPIB interface and assisted by a computer program generating also the
waveforms needed.

In order to obtain a set/sequence of FORCs, a series of triangular voltage
pulses with increasing amplitude [see Figs. \ref{Fig_2}(a$_{1}$) and (a$_{2}$%
)] is generated and applied to the solenoid via the high speed amplifier.
The frequency of an ideal triangular wave for each FORC can be defined as $%
f=\upsilon _{H}\left/ 2\left( H_{\mathrm{sat}}-H_{\mathrm{reversal}}\right)
\right. $, where $H_{\mathrm{sat}}$ and\ $H_{\mathrm{reversal}}$ are the
saturation and reversal field, respectively, while $\upsilon _{H}$ is the
applied field sweep rate. The saturation field was chosen so that the
hysteresis properties of the magnetic sample no longer change when larger
field amplitudes are applied. Frequency $f$ decreases as\ $H_{\mathrm{%
reversal}}$ decreases from $+H_{\mathrm{sat}}$ to $-H_{\mathrm{sat}}$. Using
a triangular signal we obtain a linear variation of the applied field as a
function of time (i.e., a constant $\upsilon _{H}$ on each FORC), what is
useful in the analysis of the kinetic effects, which are also time
dependent. Of course, a FORC sequence with a constant frequency $f$ can be
used, but this complicates the results' analysis because in this case the
sweep rate will increase as the reversal field decreases. In addition,
maintaining a constant frequency and increasing/decreasing the signal's
amplitude, sweep rate will increase/decrease as well. Moreover, using a
sinusoidal signal would give rise to strong nonlinear phenomena.

The frequency spectrum of a triangular wave contains only odd harmonics
which amplitude decreases proportional to the inverse square of the harmonic
number. Critical to the production of the fast transitions in the applied
magnetic field $H_{\mathrm{applied}}$ is the high speed, broad band linear
amplifier, able to amplify as many harmonics of $f$. Leaving out the high
frequency components has an effect of smoothing the $H_{\mathrm{applied}}$ 
\textit{versus} time curve. The frequency response is limited also by the
ratio of the inductance/reactance to the resistance of solenoid and $R_{H}$
assembly, ratio that should be as small as possible (e.g., using the lowest
possible number of turns in the solenoid construction or decreasing their
diameter, while the maximum value of $R_{H}$ is limited by the amplifier's
maximum output voltage). The inductance should be as small as not to block
the high frequencies. Additionally, there is always non-zero capacitance
between any two conductors which can be significant at high frequencies with
closely spaced conductors. At low frequencies parasitic capacitance can
usually be ignored, but in high frequency circuits it can be a problem
because the parasitic capacitance will resonate with the inductance at some
high frequency making a self-resonant inductor, leading to parasitic
oscillations. Above this self-resonant frequency the inductor actually has
capacitive reactance, as well. Consequently, in designing a dynamic FORC
magnetometer is important to use coils (both the solenoid that creates the
excitation field and the detection coils) with a self-resonant frequency
higher than the frequency of the applied signal. The stray inductance and
capacitance can be minimized for example by careful separation of turns,
wires and components and by keeping the leads very short. Frequency
dependent impedances are sources of systematic errors that should be
considered in the analysis of experimental dynamic FORCs. The frequency
dependence of the impedance of our coils was measured using a precision LCR
meter and we have found the self-resonant frequency higher than $2\,\mathrm{%
MHz}$. Due to increased inductive reactance with increasing frequency as $%
\upsilon _{H}$ increases (i.e., the high frequency content of $H_{\mathrm{%
applied}}$ increases) a rounding of the tips of the triangular signals is
obtained, and consequently the instantaneous field sweep rate around the
reversal field is variable and smaller. Accordingly, when the decreasing
applied field approaches the reversal field $H_{\mathrm{reversal}}$ the
paths taken by the magnetic moment differ from the major hysteresis loop -
see Figs. \ref{Fig_3}(a$_{1}$-d$_{1}$) where in the dFORCs the gray curves
represent the portions with a decreasing field. This is another source of
systematic errors in obtaining high sweep rate dynamic FORCs, but only
affecting the results around $H_{\mathrm{applied}}=H_{\mathrm{reversal}}$.
Throughout this paper we have calculated the experimental sweep rate on each
FORC as the slope of a linear fit both on the decreasing as well as
increasing part of the applied field, after removal of the first and last
quarter of each part in order to eliminate the effect of rounded tips. As a
result of solenoid's inductance, the current through it lags behind the
applied voltage, and accordingly behind the waveform generator's
synchronization signal, by a phase angle depending on inductance and $%
\upsilon _{H}$, and therefore the acquired signals by the oscilloscope start
with a portion that actually belong to the last FORC. This should be
considered in the subsequent numerical processing of data.

Another problem related to signal generation for the excitation field is
related to how the waveform generator produces a piece-wise linear function
(as the signal that we need) from a finite number of discrete points who are
sent by the user from the remote interface to generator. The staircase
approximation which is used by our generator leads to unwanted oscillations
induced in the detection coils, and this effect is more important at low
frequencies. The width of the steps in the staircase approximation can be
reduced by increasing the number of points sent from the remote interface,
the maximum number of points that can be sent being a characteristic of each
generator (in our case $262,114$). This maximum number limits the number of
FORCs that can be generated in a measurement, i.e., limits the value of $%
\Delta H_{\mathrm{reversal}}$ increment from FORC diagrams. Initially we
have generated a waveform so that $H_{\mathrm{reversal}}$ starts from $+H_{%
\mathrm{sat}}$ and decreases up to $-H_{\mathrm{sat}}$, so that the entire
hysteresis loop is covered, then we have made a \textquotedblleft
zoom\textquotedblright\ generating only the FORCs that are distinct from one
another and distinct from the ascending branch of the hysteresis loop (see
Fig. \ref{Fig_2}). In this way the\ $\Delta H_{\mathrm{reversal}}$ increment
decreases and FORC diagrams with a better resolution are obtained.

From Figs. \ref{Fig_2}(b$_{1}$) and (b$_{2}$) we observe that the induced
signal in the detection coils exhibits rapid variations with sharp peaks
which imply high frequencies, and to capture these features we need short
sampling in the acquisition process. High sampling frequency is essential
when investigating dynamic magnetic properties, because low sample rate
distorts the original signal, while high sample rate can properly reproduce
the real signal. For a fast measurement of FORC diagrams, with a good
resolution, we have used the high speed acquisition systems of a DPO7254
Tektronix digital oscilloscope. If a data acquisition card with a lower
sample rate is available, then the signal presented in Fig. \ref{Fig_2}(a$%
_{1}$) should be broken up into smaller pieces, with smaller numbers of
FORCs, measure the corresponding FORCs, and finally the pieces are added
back together. Generation of each piece should be done so as to maintain the
same field sweep rate.

The raw experimental data were first smoothed using a numerical FFT (fast
Fourier transform) filter, the induced signal was numerically integrated to
obtain the dynamic FORCs, and each FORC was numerically corrected for drift
using the saturation values. In order to reduce the errors due to the
miss-compensation of the detection coils, the signal induced in the
detection coils with no sample in was acquired, and afterwards it was
subtracted from the signal with sample. Because the mixed derivative from
the FORC diagram's definition significantly amplifies the measurement noise,
the sensitivity of the measurement technique is important, as is the
smoothing and numerical differentiation method applied to the raw data.

\section{Results}

\begin{figure*}[tbp]
\includegraphics[width=173mm,keepaspectratio=true]{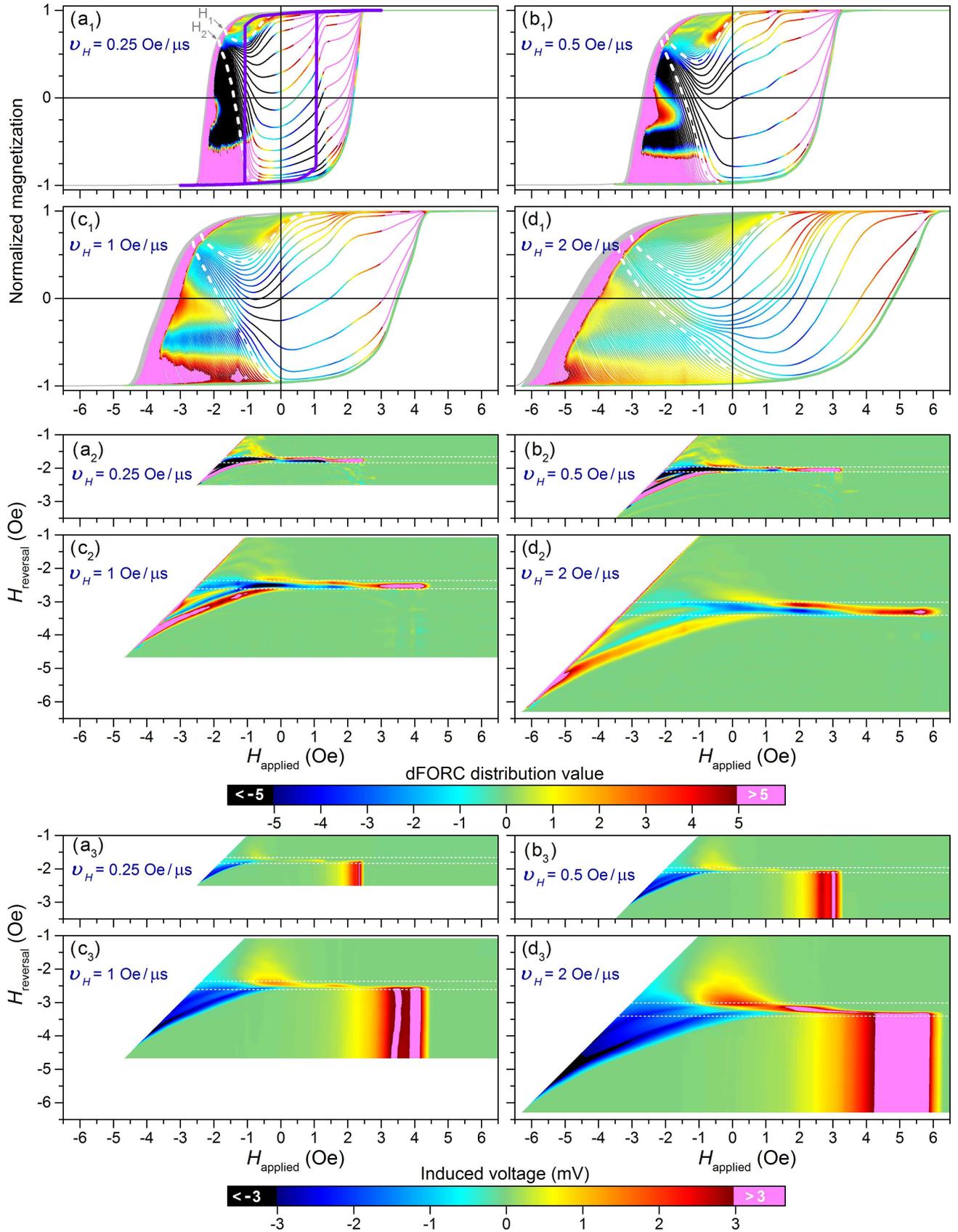}
\caption{Experimental normalized dFORCs (a$_{1}$-d$_{1}$), dFORC diagrams (a$%
_{2}$-d$_{2}$), and induced voltage (which is proportional with the
differential susceptibility $\protect\chi _{\mathrm{differerntial}}$) in the
detection coils (a$_{3}$-d$_{3}$) for different values of the applied field
sweep rate $\protect\upsilon _{H}$. The thick curve from (a$_{1}$)
represents the quasistatic MHL measured with a VSM. The gray curves from
dFORCs represent the portions corresponding to a decreasing field. Due to
the dynamic effects the magnetic moment continues to decrease even after $H_{%
\mathrm{applied}}$ starts to increase. The dotted white curves delimit three
regions: (i) $H_{1}<H_{\mathrm{reversal}}<+H_{\mathrm{sat}}$: the
corresponding dFORCs either coincide with the descending branch of dMHL, or
rapidly converge towards it, these dFORCs mainly covering the upper-left
corner of dMHLs' interior; (ii) $H_{2}<H_{\mathrm{reversal}}<H_{1}$: the
corresponding dFORCs cover the most of dMHLs' interior; and (iii) $-H_{%
\mathrm{sat}}<H_{\mathrm{reversal}}<H_{2}$: the corresponding dFORCs
converge towards the ascending branch of dMHL, these dFORCs mainly covering
the lower-left corner of dMHLs' interior.}
\label{Fig_3}
\end{figure*}

\begin{figure}[tbp]
\includegraphics[width=75mm,keepaspectratio=true]{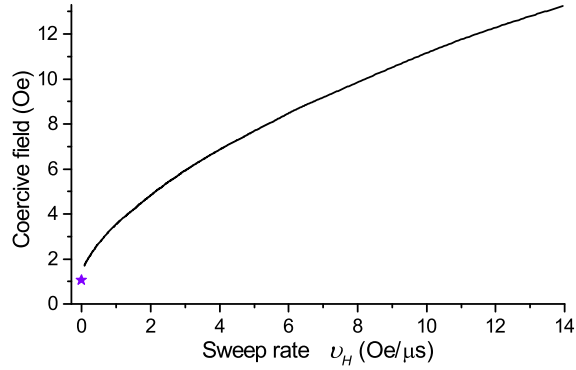}
\caption{Coercive field measured on dMHLs as a function of sweep rate. The
star marks the coercive field measured on the quasistatic MHL.}
\label{Fig_4}
\end{figure}

\begin{figure*}[tbp]
\includegraphics[width=180mm,keepaspectratio=true]{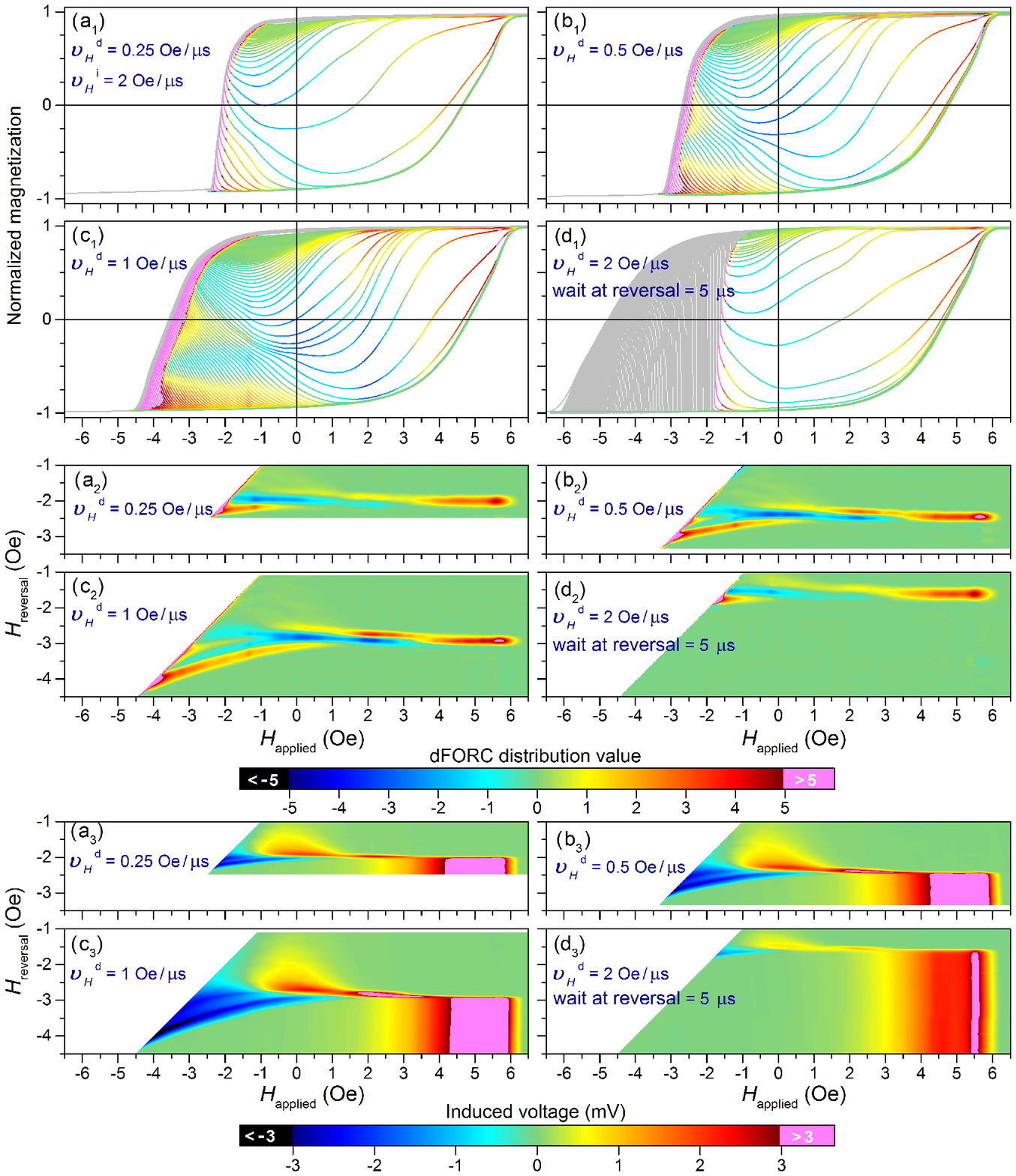}
\caption{Similar with Fig. \protect\ref{Fig_3}, but for different values of
the decreasing applied field sweep rate (a-c) and for a waiting time of $5\,%
\mathrm{\protect\mu s}$ at the reversal points (d), while the increasing
applied field sweep rate has the same value $\protect\upsilon _{H}^{i}=2\,%
\mathrm{Oe}\left/ \mathrm{s}\right.$ for all cases [value corresponding to
Figs. \protect\ref{Fig_3}(d)].}
\label{Fig_5}
\end{figure*}

All the experiments were performed at room temperature, on a $\mathrm{Fe}%
_{75}\mathrm{B}_{15}\mathrm{Si}_{10}$ $7\,\mathrm{mm}$ long amorphous
glass-coated microwire with the diameter of the metallic nucleus of $8.5\,%
\mathrm{\mu m}$ and the glass coating thickness of $6.5\,\mathrm{\mu m}$.
The magnetic domain structure of these wires typically consists of an inner
core axially magnetized, occupying most of the microwire's volume,
surrounded by an outer shell (OS) containing domains with radial easy axes
of magnetization (see, e.g., Ref. \onlinecite{Vazquez 2015}). Such materials
are a cheaper alternative to nanowires prepared using various lithographic
techniques, while their magnetic properties can be tailored in a wide range
through composition or sample dimensions.\cite{Vazquez 2015, Zhukov 2009}
The external magnetic field was applied parallel to the wire axis. During
our experiments that measured the dynamic MHLs and FORCs the sample was
entirely placed in one of the detection coils, so that the induced signal is
due to the time variation of the microwire's total magnetic moment.
Throughout this paper the magnetization curves are normalized to saturation.

The quasistatic major hysteresis loop (MHL) measured with a vibrating sample
magnetometer (VSM), and represented with a thick curve in Fig. \ref{Fig_3} (a%
$_{1}$), indicates that the uniaxial magnetic anisotropy of the magnetic
wire determines virtually a bistable magnetic behavior between two stable
axially magnetized states. However, the loop is not perfectly rectangular,
but is slightly rounded near the switching/coercive field. As the applied
field is reduced from its maximum value, there is a first region in which
the normalized magnetization $m$ decreases slowly from $1$ to about $0.94$.
This region can be mainly attributed to the magnetization rotation in the
outer shell, from the axial direction to perpendicular directions, at
remanence the total magnetization of the outer shell being equal to zero.
The magnetization of the entire wire at remanence is $0.95$. Concurrently
begins the formation of closure domain structures at the ends of the wire,
where the spins re-distribute to reduce the magnetostatic/stray energy.
Further decrease of the applied field results in a much quicker decrease of
the magnetization to about $0.8$, and it can be linked to an angular
magnetization reorientation and/or by an enlargement of the closure
structures. The third region of the quasistatic MHL is the one step reversal
of the magnetization, also known as the large Barkhausen effect.

The rectangular-shaped hysteresis loop indicates that after the formation of
the closure domains, the magnetization reversal process between the two
stable magnetic configurations occurs by depinning and propagation of a
domain wall (DW) inside the single domain inner core, from the closure
structure at one end of the wire (see, e.g., Refs. 
\onlinecite{Zhukov 2009, Ipatov JAP
2009, Varga PRL 2005, Vasquez PRL 2012, Ye JAP 2013, Zhukova JAC 2014}). The
switching is determined by the field needed to nucleate the reverse domain.
With increasing wire length, increases the probability of nucleating some
new additional reversed domains and new domain walls inside the wire (see,
e.g., Refs. \onlinecite{Ipatov JAP 2009, Corodeanu RSI 2011, Ovari 2012}).
The control of the magnetization reversal process and particularly of the
domain wall's motion is essential for the development of novel applications
using magnetic wires. Due to the sudden decrease in magnetization at the
switching point, the interior part of the quasistatic MHL can not be covered
by quasistatic FORCs.

Since the microwires show in many cases inhomogeneities in their properties
(geometry imperfections/defects, surface roughness produced during the
fabrication process, impurities, or compositional irregularities ascribed to
the amorphous structure), several pieces of wires were measured, all pieces
having the same length, but being cut from different portions of an initial
long wire. The quasistatic MHLs are similar, rectangular-shaped, only small
variations of the switching field from sample to sample. However, even if
the dynamic major hysteressis loops are also similar, the dynamic FORCs have
features that are distinctive for each sample. The dFORCs have main
characteristics which are common for all the investigated samples, but they
also have fine distinguishing features that differ from sample to sample.
The dFORCs have proven to be a much more sensitive tool in regard with the
hysteresis loop, revealing various distinctive features of the magnetization
switching, revealing the contribution of local inhomogeneities on the
peculiarities of DW propagation, and therefore a useful tool for
investigating the fine details of the magnetization dynamics and reversal.
This method can be used also as a fast, nondestructive method for
investigating the samples' homogeneity.

In Fig. \ref{Fig_3} are presented the experimental normalized dynamic FORCs
(a$_{1}$-d$_{1}$) for one of the samples, the corresponding dynamic FORC
diagrams (a$_{2}$-d$_{2}$), and the induced voltage measured across the
detection coils (a$_{3}$-d$_{3}$), for different values of the applied field
sweep rate $\upsilon _{H}$, the sweep rate of the decreasing field being
equal to that of the increasing field. The gray curves from dFORCs represent
the portions corresponding to a decreasing field. One observes that both the
dMHL and dFORCs as well dFORC diagrams are dependent on the sweep rate.

We observe that an increase in sweep rate $\upsilon _{H}$ increases the
slope of the vertical regions of the dynamic hysteresis loops, with respect
to the quasistatic hysteresis. This phenomenon can be attributed to the
counterbalance between the time required for the $+H_{\mathrm{sat}}$ to $-H_{%
\mathrm{sat}}$ variation, and the switching time related to the time needed
for DW propagation along the whole wire. This kind of hysteresis is often
referred to as rate-dependent hysteresis. The coercive field (defined as the
field at which the projection of the magnetization on the applied field
direction is zero) increases as $\upsilon _{H}$ increases (see also Fig. \ref%
{Fig_4}).

In Figs. \ref{Fig_3} one observes that due to the dynamic effects the
magnetic moment continues to decrease even after the applied magnetic field
starts to increase. The last FORC (with the smallest $H_{\mathrm{reversal}%
}<0 $ value) gives in fact the dynamic major hysteresis loop (dMHL). We note
that the frequency $f_{\mathrm{MHL}}$ is provided in Fig. \ref{Fig_2} only
as guidance, the sweep rate $\upsilon _{H}$ being in fact the parameter
which properly describes dMHLs, dFORCs, and dFORC diagrams, because the
kinetic effects depend on $\upsilon _{H}$. Due to the rounding of the tips
of the triangular applied field, the magnetization curves differ from the
dynamic hysteresis loop when the decreasing applied field approaches the
reversal field [see the gray portions in dFORCs in Figs. \ref{Fig_3}(a$_{1}$%
-d$_{1}$)]. Owing to the dynamic/kinetic effects the magnetization continues
to decrease even after the applied magnetic field starts to increase, after
the reversal points, i.e., there is a lag of the output (magnetization) with
respect to the input (excitation applied magnetic field). These portions
correspond to a negative induced voltage across the detection coils. The lag
between magnetization and the applied field is more prominent as $\upsilon
_{H}$ increases. Usually a lag between input and output is associated with a
complex variable, i.e., with a real and imaginary component, the imaginary
component indicating the dissipative processes in the sample. Hereby we can
define the dynamic complex FORC (dcFORC).

We can define three regions in the dFORCs from Figs. \ref{Fig_3} (a$_{1}$-d$%
_{1}$): (i) $H_{1}<H_{\mathrm{reversal}}<+H_{\mathrm{sat}}$ for which the
corresponding dFORCs either coincide with the descending branch of dMHL, or
rapidly converge towards it, these dFORCs mainly covering the upper-left
corner of dMHLs' interior; (ii) $H_{2}<H_{\mathrm{reversal}}<H_{1}$ for
which the corresponding dFORCs cover the most of dMHLs' interior; and (iii) $%
-H_{\mathrm{sat}}<H_{\mathrm{reversal}}<H_{2}$ for which the corresponding
dFORCs converge towards the ascending branch of dMHL, these dFORCs mainly
covering the lower-left corner of dMHLs' interior. The borders between these
regions are marked with dotted white curves in Fig. \ref{Fig_3}.

The dFORCs from the first region can be associated with almost reversible
magnetization processes related to the generation of the closure domain
structure. As the field $H$ is not sufficient to generate large scale
propagation of the walls on these FORCs, what one observes are the processes
of wall nucleation when the field is negative and with increased absolute
value and as soon as the field is again increasing on each FORC the wall
disappears as the positive saturation is attained. When the negative field
is sufficient to start a considerable propagation of the wall towards the
negative saturation the FORCs are significantly different. This is the
second region on the set of FORCs and the richest in information due to the
fact that these FORCs are actually covering a major part of the dMHLs'
interior.

One observes that these dFORCs have several inflection points which
demarcate portions with different slopes on a given dFORC, the slope giving
the differential susceptibility $\chi _{\mathrm{differerntial}}$. The
inflection points that are near the ascending branch of the quasistatic
hysteresis loop in Fig. \ref{Fig_3}(a$_{1}$) separate portions with very
different slopes, which leads to the idea of switching of two different
magnetic \textquotedblleft entities,\textquotedblright\ such as the inner
core and the outer shell of the microwire, the coercive field or the end of
the switching of the outer shell being given by the inflection points. We
note that these switching events are not very visible on dMHL. The magnetic
field corresponding to the inflection points increases with increasing sweep
rate. The portions of dFORCs located to the left of these points give rise
to a pair of negative and positive peaks on the diagram, while the portions
located to the right give rise to a single positive peak. The peaks with $H_{%
\mathrm{applied}}>0$ in the dFORC diagrams are mainly given by the dFORCs
from the second region, but because the $\left[ H_{1},H_{2}\right] $
interval is rather narrow, the FORC diagram features are constraint along a
horizontal line, parallel with the $H_{\mathrm{applied}}$ axis, in contrast
with most of the results presented in the scientific literature (where
usually the FORC diagram's peaks lie along the $H_{\mathrm{applied}}=H_{%
\mathrm{reversal}}$ and/or $H_{\mathrm{applied}}=-H_{\mathrm{reversal}}$
diagonals). While both fields $H_{1}$ and $H_{2}$ indicating the second
region on the FORC set are gradually shifting towards more negative fields
with increasing sweep rate, the width of the $\left[ H_{1},H_{2}\right] $
interval increases in the same time.

For the dFORCs from the third region, the magnetization's reversal started
on the descending branch of dMHL can not be stopped by the increasing
applied field, the magnetic wire reaching the negative saturated state, and
then going back to the positive saturated state. The ascending part of these
dFORCs coincide with the ascending branch of dMHLs (the induced voltage in
the detection coils being the same), giving no peaks in the corresponding
regions of the dFORC diagrams. The descending part of these dFORCs gives
rise to several positive and negative peaks in diagrams, peaks that start
along the $H_{\mathrm{applied}}=-H_{\mathrm{reversal}}$ diagonal (which in
the Preisach model corresponds to the coercive field axis) and converge
towards the \textquotedblleft horizontal\textquotedblright\ peaks generated
by the dFORCs from the second region. These peaks displaces down on the $H_{%
\mathrm{reversal}}$ axis with increasing sweep rate. All peaks broaden as $%
\upsilon _{H}$ increases.

In Figs. \ref{Fig_3}(a$_{3}$-d$_{3}$) the induced voltage in the detection
coils is presented in the FORC coordinates, i.e., as a function of $H_{%
\mathrm{applied}}$ and $H_{\mathrm{reversal}}$. The induced voltage is
proportional with the rate of change of the sample's magnetization, i.e.,
with the derivative $dm\left/ dt\right. $ of the magnetization with respect
to time. Taking into account that $dm\left/ dt\right. =\left( dm\left/ dH_{%
\mathrm{applied}}\right. \right) \left( dH_{\mathrm{applied}}\left/
dt\right. \right) =\upsilon _{H}\left( dm\left/ dH_{\mathrm{applied}}\right.
\right) =\upsilon _{H}\,\chi _{\mathrm{differerntial}}$, the induced voltage
would be proportional with the derivative of the magnetization with respect
to applied field, i.e, with the differential susceptibility $\chi _{\mathrm{%
differerntial}}$, for a perfect triangular applied field. However, as we
have shown in the experimental setup description, due to the rounding of the
tips of the triangular signals from Fig. \ref{Fig_2}(a$_{1}$), the
instantaneous field sweep rate around the reversal field is variable.
Nevertheless, excepting the region close to $H_{\mathrm{applied}}=H_{\mathrm{%
reversal}}$, for a given sweep rate $\upsilon _{H}$, the induced voltage
from Figs. \ref{Fig_3}(a$_{3}$-d$_{3}$) is proportional with the
differential susceptibility.

The results we have discussed already are rising the problem that even if
the FORCs are measured only when the field is increasing, the results might
be dependent also on the magnetization processes during the preparatory
steps performed in experiment to generate the initial points on each FORC,
when the field is actually decreasing. We have designed a few experiments to
evidence this dependency.

In Figs. \ref{Fig_5}(a-c) the sweep rate of the increasing field is held at
the value from Figs. \ref{Fig_3}(d), while the decreasing sweep rate is
decreased. The use of two different sweep rates leads to an asymmetry of the
dynamic hysteresis loop. As $\upsilon _{H\mathrm{,\,decrease}}$ decreases
(i.e., the decreasing branch of the hysteresis approaches the quasi-static
case) the negative part of the induced voltage is smaller than the positive
part. The quasi-static decreasing branch of MHL can be reached also by
waiting at the reversal points, so that the sample's magnetization relaxes
in a constant field. Systematic experiments have shown that this waiting has
a similar effect with that from Figs. \ref{Fig_5}(a-c), the dFORC diagram
obtained with a waiting time at the reversal points is similar with the
dFORC diagram from Fig. \ref{Fig_5}(a$_{2}$). The decay of the magnetization
during the waiting time is represented with gray curves in dFORCs from Figs. %
\ref{Fig_5}(a$_{1}$-d$_{1}$). As the waiting time increases, the end points
of these gray curves converge towards the quasistatic MHL.

Related to this type of experiment, we should mention that most VSM
producers (e.g. PMC MicroMag VSM) have provided the possibility of setting a
waiting time in the reversal point in the FORC experimental procedure.
However, as the FORC procedure is not standardized yet from this point of
view, various groups are treating differently this aspect and when the data
are processed (e.g. with FORCinel \cite{FORCinel}) the first point on each
FORC is treated as an \textquotedblleft uncertain\textquotedblright\ data
and can be removed even if we can track the physical cause of the apparently
erratic behavior in the reversal point. The experimental systematic study
presented in this article could provide a guide for the correct treatment of
data when the kinetic effects are essential in magnetic systems (see also
Ref. \onlinecite{Rotaru EPJB 2011} for a study of the kinetic effect in the
FORC reversal points for the temperature hysteresis in spin-transition
materials).

\section{Model}

In order to develop a model able to describe the magnetization process of
our sample, we have started from the results of the studies showing that the
domain structure of the amorphous magnetic wires typically consists of an
inner core, occupying most of the microwire's volume, and an outer shell
with various orientations of their easy axes of magnetization (see Refs. %
\onlinecite{Zhukov 2009, ch.12 Vazquez 2015}, and references within it). Due
to the absence of the long range ordering of atoms, amorphous microwires do
not display magnetocrystalline anisotropy, but the cylindrical geometry of
the wire gives rise to significant uniaxial shape anisotropy with a
longitudinal magnetization easy axis. In Fe-based wires the magnetoelastic
anisotropy reinforces the shape anisotropy leading to a quite large single
domain core axially magnetized. Studies using the magneto-optical Kerr
effect (MOKE) have shown that in some cases the surface magnetization
reverses at the same longitudinal applied field as the main axial domain,
which suggests that the magnetization at the surface is either part of the
axially magnetized domain, or that it contains a significant axial
contribution.\cite{ch.12 Vazquez 2015, Zhukov 2009} Due to this domain
structure, the magnetization reversal of amorphous magnetic microwires with
axial anisotropy under homogeneous longitudinal field takes place by the
depinning and propagation of a domain wall from a complex closure domain
structure near the microwire's end.

\begin{figure}[tbp]
\includegraphics[width=85mm,keepaspectratio=true]{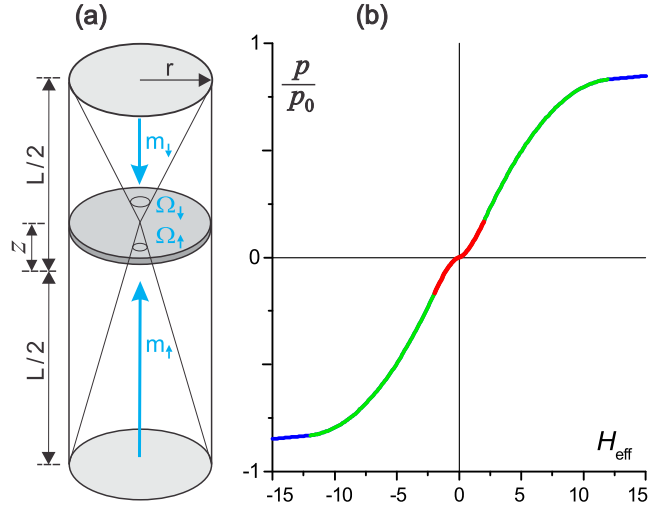}
\caption{(a) Proposed magnetic domain structure of the wire, where a
propagating domain wall separates two domains with antiparallel oriented
magnetizations. (b) Dependence of the pressure term as a function of the
effective field. The three sub-domains described in text are emphasized with
different colors. In the case of a uniform motion of the wall, its constant
velocity is proportional with the pressure term.}
\label{Fig_6}
\end{figure}

To model the experimental results in a simple way, we assume that the
magnetization processes in these cylindrical wires consist in the
displacement of a $180^{\circ }$ domain wall along the entire length of the
sample, the wall remaining plane and perpendicular to the cylinder's axis
during its motion, as shown Fig. \ref{Fig_6}(a). We describe the wall as a
two-dimensional geometrical surface, which means that we are neglecting the
internal wall structure and the wall thickness. The equation of motion of
the domain wall is taken on an analogy between the dynamics of the domain
wall and the dynamics of a mechanical viscously damped forced oscillator
(see, e.g., Refs. 
\onlinecite{Chikazumi, Cullity, ch.12 Vazquez 2015, Zhukov
MSE 1997, Varga PRL 2005, Vasquez PRL 2012})%
\begin{equation}
\sigma \frac{d^{2}z}{dt^{2}}+\beta \frac{dz}{dt}+\alpha z=p,
\label{dw_motion_0}
\end{equation}%
where $z$ describes the wall's position [see Fig. \ref{Fig_6}(a)], $\sigma $
is the equivalent mass of the wall per unit area, $\beta $ is the viscous
damping parameter, $\alpha $ is the restoring coefficient, and the term $p$
on the right hand side of the equation represents the pressure acting on the
wall, created by the applied field.

When a domain wall moves, a precessional motion of the moments associated
with the wall occurs, and as a result, the wall behaves like it possesses a
mass, despite the absence of any actual mass displacement. If in the
classical Newton's mechanics the inertial mass of an object (particle) is
determined by its resistance to acceleration due to action of an external
force, the inertia of a domain wall reflects an influence of deformations of
a moving domain wall on its energy (i.e., how this energy depends on
velocity). \cite{Doring 1948} A propagating domain wall could continue to
propagate, even in the absence of an external magnetic field. The inertia of
the wall, or the resistance of the spins to sudden rotation, is not usually
important except at very high frequencies.

The second term in the above equation represents a resistance to motion
which is proportional to the velocity, and stands for the dissipated energy
when the domain wall moves. Three causes of viscous damping have been
identified in the scientific literature: production of eddy currents (and
consequent Joule heating) around the moving wall, an intrinsic or relaxation
effect (arising from the fact that the domain wall velocity is restricted by
intrinsic damping of magnetic moments of the domain wall), and a structural
relaxation (arising from the interaction of atomic defects with the domain
wall). The eddy currents, in turn, produce a magnetic field that acts to
oppose the wall motion.

The third term $\alpha z$ represents the restoring force due to crystal
imperfections such as microstress or inclusions, $\alpha $ being related to
the shape of the potential energy minimum in which the wall is located. The
value of $\alpha $ determines the field required to move the wall out of the
energy minimum, and the ensemble of $\alpha $ values for the entirely sample
determines the coercive field required for extensive wall movement. When the
wall moves with the velocity $\upsilon $, the total magnetic moment of the
two domains adjacent to the wall changes at the rate $2M_{s}\upsilon $ per
unit area of wall, where $M_{s}$ is the wire's saturation magnetization.

When a magnetic field $H$ is applied parallel to the direction of
magnetization of a domain on one side of the wall, the density energy of the
system decreases by $2M_{s}H\Delta z$ when the wall is displaced by $\Delta
z $, i.e., the term $2M_{s}H$ acts as a pressure on the wall surface.
Accordingly, the expression%
\begin{equation}
p=2M_{s}H  \label{p}
\end{equation}%
is commonly used in the scientific literature for the pressure term in the
right hand side of the Eq. (\ref{dw_motion_0}).

Once the wall propagates at the constant velocity $\upsilon _{0}$ under the
constant field $H$, then the first term of Eq. (\ref{dw_motion_0}) drops out
because the acceleration is zero, while the third term should be modified
because the wall is moving large distances and is not affected by the value
of $\alpha $ at one particular energy minimum. Instead, the third term will
represent the average resistance to wall motion. In this case from Eqs. (\ref%
{dw_motion_0}) and (\ref{p}) follows that:%
\begin{equation*}
\upsilon _{0}=\frac{2M_{s}}{\beta }\left( H-H_{c}\right) ,
\end{equation*}%
where $H_{c}$\ is the field which should be exceeded before extensive wall
motion can occur, and it is approximately equal to the coercive field (or
switching field of the axial domain). This linear dependence of the wall
velocity on the applied magnetic field is reported in many works, starting
with the pioneering work of Sixtus and Tonks in 1931.\cite{Sixtus Tonks}

However, the velocity of the DW propagation does not increase infinitely,
but saturates at some field, the so called Walker limit field.\cite{Walker
JAP 1974} Above this limit field, the velocity may fluctuate or even
oscillates, due to the fact that the spin structure inside the DW is varied
by the high magnitude field. On the other hand, when the applied field is
smaller than the coercive field the domain wall moves by successive
thermally activated jumps. In this low field range it was experimentally
found that for thin films the wall's velocity increases exponentially with
the field.\cite{Rio-Labrune IEEE 1987} In Ref. 
\onlinecite{Zhukova AIP-CP
2008} is reported that for $\mathrm{Fe}_{69}\mathrm{B}_{15}\mathrm{Si}_{10}%
\mathrm{C}_{6}$ microwires the dependence $\upsilon _{0}\left( H\right) $ of
the velocity on the applied magnetic field is essentially not linear,
exhibiting two or three different regimes (depending on the diameter of the
wire), with significantly higher domain wall mobility $d\upsilon _{0}\left/
dH\right. $ at low field limit. This non linearity was attributed to the
change of the DW structure with changing of the applied magnetic field.

This non-linear dependence of the domain wall's velocity on a constant
applied magnetic field is not retrieved from Eqs. (\ref{dw_motion_0}) and (%
\ref{p}). Therefore we propose to introduce the non-linear dependence of the
wall velocity in the pressure term from the right hand side of Eq. (\ref%
{dw_motion_0}). In the case of a uniform motion of the wall, its constant
velocity is proportional to the pressure term. This proportionality suggests
the possibility to evaluate the pressure term from the experimental
dependence $\upsilon _{0}\left( H\right) $ of the wall velocity as function
of the constant field.

In order to find the $\upsilon _{0}\left( H\right) $ dependence for our
sample, we have used the same experimental setup as for dFORC measurements,
but we have applied a constant magnetic field preceded by a short magnetic
field pulse able to initiate the DW depinning. After the initial pulse the
wall propagates along the entire microwire under the effect of the constant
applied magnetic field. The induced voltage in the detection coils is
proportional with the time derivative of the magnetization $dm\left/
dt\right. $, i.e., with the wall's velocity. The procedure was repeated for
many values of the constant field to obtain the $\upsilon _{0}\left(
H\right) $ dependence. After overcoming the energy barrier corresponding to
the DW nucleation and depinning, the DW propagates even in a field smaller
than the switching field.

To approximate the experimental data analytically we have used the
cumulative distribution function (CDF) of a generalized trapezoidal
distribution \cite{Dorp Metrica 2003}, because it seems to be appropriate
for modeling the duration and the form of a phenomenon which may be
represented by two or three stages [see Fig. \ref{Fig_6}(b)]. In principle,
the fitting function is a piece-wise function obtained by concatenating in a
continuous manner a power function with a $n_{1}>0$ exponent, a quadratic
polynomial, and a power function with a $n_{3}>0$ exponent, the length of
each sub-domain being adaptable. Its flexibility allows to appropriately
mimic a great variety of behaviors. In our simulations $n_{1}=1.5$ and $%
n_{3}=3$, respectively.

The effective magnetic field $H_{\mathrm{effective}}=H+H_{i}$ acting on DW,
incorporates the applied field and the internal field $H_{i}$. The internal
field may contain the demagnetizing field, the anisotropy field, and any
other internal field. In order to keep our model simple, in this paper we
take into account only the demagnetizing field computed in the middle of the
wall, arising from the magnetic uncompensated polarization charges which lie
at each end of the cylinder:

\begin{eqnarray}
H_{d} &=&M_{s}\left( \Omega _{\uparrow }-\Omega _{\downarrow }\right)
\label{Hd} \\
&=&2\pi M_{s}\left( \frac{L\left/ 2\right. +z}{\sqrt{r^{2}+\left( L\left/
2\right. +z\right) ^{2}}}-\frac{L\left/ 2\right. -z}{\sqrt{r^{2}+\left(
L\left/ 2\right. -z\right) ^{2}}}\right) ,  \notag
\end{eqnarray}%
where $\Omega _{\uparrow }$ and $\Omega _{\downarrow }$ are the solid angles
subtended by the cylinder's bases at the middle of the wall, while $r$ and $%
L $ are the cylinder's radius and length, respectively [see Fig. \ref{Fig_6}%
(a)]. The demagnetizing field is equal to zero when the wall is in the
middle of the wire and its absolue value increases to $2\pi M_{s}L\left/ 
\sqrt{r^{2}+L^{2}}\right. $ as the wall moves to one extremity. Accordingly,
the effective field acting on the DW is a function on the wall's position.
The magnetization reversal process depends on the geometry of the magnetic
wire through the corresponding variations of demagnetizing field. Of course,
the demagnetizing field could be calculated as the average over the wall's
surface. In order to take into account the length of the closure domains, or
the critical displacement of the DW to be depinned, in simulations we have
taken the maximum value of the DW's displacement as being $z_{max}\left/
\left( L\left/ 2\right. \right) \right. =0.8$, value given by the normalized
magnetization just before the switching on the experimental quasistatic MHL
[see Fig. \ref{Fig_3}(a$_{1}$]. The demagnetizing field is oriented so as to
saturate the magnetic wire, i.e., the demagnetizing field acts as a pining
field of the wall at the ends of the wire. If the sample is saturated, then
only a magnetic field with a greater value than that of the demagnetizing
field, and acting in opposite direction can start the wall's motion.
However, using the saturation magnetization's value obtained from the
experimental quasistatic MHL, would obtain a switching/pinning field value
much greater than the experimental value. This discrepancy is due to the
fact that in Eq. (\ref{Hd}) one considered that the magnetic charge density
at each end of the wire is equal to the saturation magnetization, i.e., it
was considered that there are no closure domain structures at the ends of
the wire. The closure domains reduce the total effective magnetic
uncompensated charge density, and its value was obtained by the experimental
switching field value's fit.

As we have shown above, the demagnetizing field given by Eq. (\ref{Hd}) is
equal to zero in the middle of the wire, otherwise tending to attract the DW
towards the nearest end of the wire, its absolute value exhibiting symmetry
with respect to the middle of the wire. However, in the experimental dFORCs
from Figs. \ref{Fig_3}(a$_{1}$-d$_{1}$) and \ref{Fig_5}(a$_{1}$-d$_{1}$) we
observe that on some dFORCs the moment begins to increase starting from
negative values, even if the applied magnetic field is still negative. If we
consider that the DW deppined from the upper end of the wire [like in Fig. %
\ref{Fig_6}(a)], then this behavior shows that the upper end of the wire
still attracts the wall, even if the wall already surpassed the middle of
the wire. This comportment can be explained by an asymmetry of the
demagnetizing field in regard with the middle of the wire. Such asymmetry
can be argued by different densities of magnetic charges on the two ends of
the wire. Indeed, if we take into account also the outer shell of the wire,
once its magnetization is reversed, the total magnetic charge on the lower
end of the wire decreases. Nevertheless, in order to keep our model as
simple as possible, in this paper we have considered a symmetric
demagnetizing field, as given by Eq. (\ref{Hd}).

As the demagnetizing field includes the field which should be exceeded
before extensive wall motion can occur, in the following simulations we have
kept only the first two terms in left hand side of Eq. (\ref{dw_motion_0}),
while the pressure function $p\left( H_{\mathrm{effective}}\right) $ from
the right hand side is given by the fitting function described above and
presented Fig. \ref{Fig_6}(b).

We have numerically solved this equation and in the following we will show
that this simple representation of the domain structure of the magnetic
wires is able to explain many peculiarities of their dynamic magnetic
behavior. The model, being intrinsically dynamic, is able to predict the
magnetization processes, as the sweep rate (or the frequency) of the applied
field increases, from the static regime to the dynamical regime. Certainly
the above model does not stand for the entire domain structure, but we can
interpret our model in terms of a pseudo-wall and two pseudo-domains which
overall describe the dynamic magnetization processes of the magnetic wires.
Even if the real magnetization process involves complex and intricate domain
structures, the above model is a way to reduce the description of the
process to few dominant degrees of freedom. We note that Eq. (\ref%
{dw_motion_0}) by itself does not contain the physics of
nucleation/annihilation of domains, nor the magnetization rotation
mechanism, i.e., this model does not deal with the approach to saturation.
To describe also these processes, Eq. (\ref{dw_motion_0}) should be coupled
with equations describing the magnetization processes of the outer shell and
of the closure domains. Nevertheless, in this paper we neglect these
processes. Obviously, full micromagnetic models can better describe the
non-uniform magnetization processes, but their numerical efficiency is
significantly lower, and this is the reason why the phenomenological models
are still used in simulations. 
\begin{figure*}[tbp]
\includegraphics[width=180mm,keepaspectratio=true]{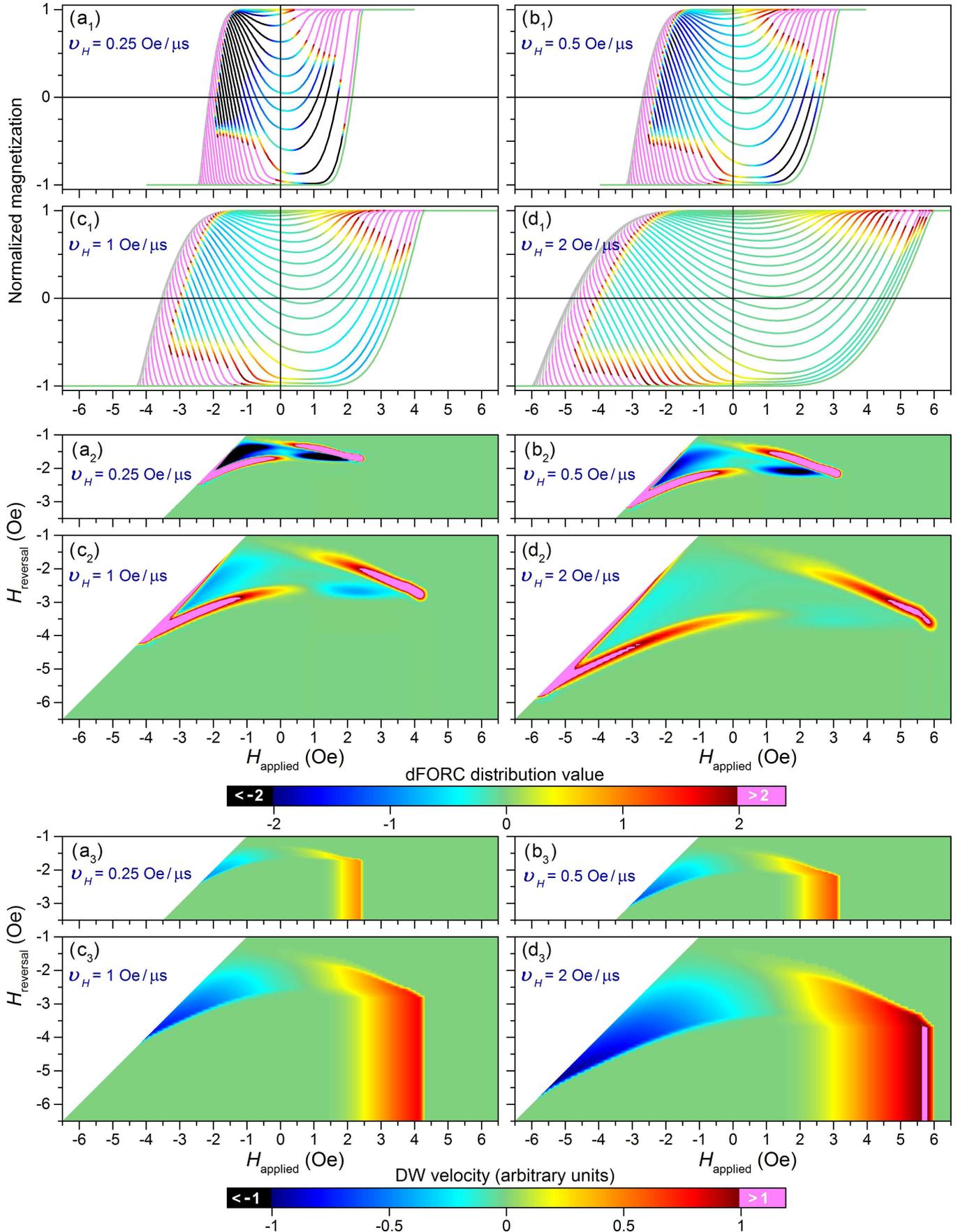}
\caption{Simulated normalized dFORCs (a$_{1}$-d$_{1}$), dFORC diagrams (a$%
_{2}$-d$_{2}$), and domain wall velocity (a$_{3}$-d$_{3}$) for different
values of the applied field sweep rate $\protect\upsilon _{H}$. The gray
curves from dFORCs represent the portions corresponding to a decreasing
field.}
\label{Fig_7}
\end{figure*}

In order to simulate the experimental data, we first numerically generated a
series of triangular magnetic field pulses with increasing amplitude, which
were then smoothed using a FFT filter, to obtain a rounding of the tips, as
it was in the case of the experiment [see Fig. \ref{Fig_2}(a)].

Starting from the positive saturation, the DW starts to move only when the
decreasing effective magnetic field become negative, i.e., when the applied
field overwhelms the threshold value of the pinning field (which, as has
been shown above, it is given by the demagnetizing field at the wire's end).
After that, the DW accelerates, the total magnetic moment of the wire being
proportional to the wall's position $z$. After the applied magnetic field
reached the appropriate reversal value and started to increase, as long as
the effective field is negative, the pressure acting on the wall surface
attempts to move it towards the lower end of the wire, and consequently the
DW continues to move accelerated toward the lower end. Accordingly, the
wire's magnetic moment continues to decrease even after the applied magnetic
field starts to increase. Furthermore, because the DW has inertia, it
continues to move down even for positive values of the pressure/effective
field. This lag depends on the mass of the wall $\sigma $, on damping $\beta 
$, and on sweep rate $\upsilon _{H}$. As $\upsilon _{H}$ increases, the lag
increases as well, and goes to zero as $\upsilon _{H}$ decreases.
Considering a linear pressure term like in Eq. (\ref{p}) and a sinusoidal
excitation, then the driving angular frequency at which the lag of
steady-state forced oscillations is maximum (i.e., at phase resonance) is $%
\omega _{0}=\beta \left/ \sigma \right. $.

Systematic simulations have shown that as the $p_{0}\left/ \sigma \right. $
ratio increases, the coercive field of dMHL decreases, while the slope $dm_{%
\mathrm{dMHL}}\left/ dH\right. $ increases. As the driving pressure $p_{0}$
exerted on DW increases (and accordingly the force pushing the wall
increases), the DW would move faster, with a higher acceleration, and a
shorter time would be required for propagation along the wire. A similar
effect has the decrease of $\sigma $ because the smaller the mass, the
larger the acceleration for a given force. As the damping $\beta $
increases, the acceleration decreases, the time required for propagation
along the wire increases, and consequently the coercive field increases,
while the slope $dm_{\mathrm{dMHL}}\left/ dH\right. $ decreases.

In our simulations $\beta \left/ \sigma \right. =1.5\,\mathrm{MHz}$ and $%
p_{0}\left/ \sigma \right. =2\times 10^{10}\,\mathrm{N}\left/ \mathrm{m}%
^{2}\right. $, values that led to a good approximation of the experimental
data, both dMHLs and dFORCs (see Fig. \ref{Fig_7}). Because Eq. (\ref%
{dw_motion_0}) does not describe the slowing and stopping of DW at the ends
of the wire, the value of the damping was gradually increased as $\left\vert
z\right\vert >0.7\,z_{\max }$ when DW moves toward the wire's end. The
double-peak structure of the experimental induced voltage [see Fig. \ref%
{Fig_2}(a)] could be explained by the existence of pinning centers along the
microwire which determine the wall's deceleration.

The dynamic susceptibility, which is the slope of the magnetization curves,
can be written as $\chi _{\mathrm{differerntial}}=dm\left/ dH_{\mathrm{%
applied}}\right. =\left( dm\left/ dt\right. \right) \left( dH_{\mathrm{%
applied}}\left/ dt\right. \right) \sim \left( dz\left/ dt\right. \right)
\left/ \upsilon _{H}\right. $, i.e., it is proportional with DW's velocity.

This fact makes the differential susceptibility diagram (first derivative of
the FORCs) probably easier to understand by the potential users of this
experimental setup. On this diagram one observes directly the wall velocity.
One sees on a typical FORC (e.g., for an initial velocity along the negative
direction -- given by a value of the reversal field) how the DW is reacting
to the field increase at a constant rate from the initial value of the
reversal field towards the positive saturation of the wire. The experimental
diagram shows at what field the wall is motionless (zero susceptibility on
the respective FORC) and how the DW is increasing its positive velocity when
the applied field is in the positive region. At very high values of the
positive applied field the DW is again motionless at the end of the wire --
again the susceptibility is zero on the FORC. As we can make experiments at
different field rates one may also observe the inertial effects on DW
movements. On these experimental data one have found in some cases a
remarkable behavior of the DW (acceleration towards positive direction in
negative applied fields) that may suggest in our opinion a brake of symmetry
in the unsaturated polarization charges at the two ends of the wire due to a
core-shell DW structure. We also would like to emphasis that most of the
experimental effects can be covered by the simple model we have proposed.

The classical FORC diagram, which is a second order derivative of the FORCs
is offering another aspect of the DW mobility. With the usual unique
sensitivity, the FORC diagram is giving details on the coercivities involved
in the wire moment switching. The region near the reversal field is quite
complicated and offers information about the process of reversal and
provides evidence of possible random sources of coercivity like small
irregularities in the wire geometrical or physical properties. The second
region which covers usually the positive field area on the diagram has a
typical structure with a doublet structure: negative for lower reversal
fields and positive at higher reversal fields. Our simple model is giving
this doublet structure but one may observe that in experiment the angle
between the positive and negative regions is significantly lower than in the
simulations. As the mentioned structure is closely related to the
magnetization processes near saturation we recognize the need for testing
more realistic terms for the approach to saturation as it is generally known
that the closure structure can be quite intricate. We also see in
experimental FORCs evidence of the influence of the shell wall which was
neglected in the model. A model with a core-shell structure for the wire
would be a logical extension of the model presented in the present article.
Also one envisage for a further study experiments with longer wires to test
the role of the closure domain wall structure.

\section{Conclusions}

In this article we have introduced a systematic study of kinetic dominated
hysteretic processes based on the FORC diagram methodology (dynamic FORC).
We propose a number of experiments able to evidence the role of kinetics in
the actual hysteretic processes observed in magnetic wires. We have
discussed in this article actually two experimental diagrams based on FORC
data: (i) the FORC susceptibility diagram is offering a direct image of the
DW velocity during the magnetization process along the different FORCs
characterized by distinct values of the reversal fields; (ii) the classical
FORC diagram calculated as the mixed second order derivative of the FORC
data as a function of the reversal and actual applied fields, that gives a
supplementary insight on the distribution of coercivities in the system due
to various physical causes. The exceptional sensitivity of the FORC diagram
method evidenced a number of typical alterations of the quasi-static FORC
distribution for a magnetic wire when the field rate is increased. In order
to provide the study with a quantitative attribute we have developed a model
able to describe rather accurately the entire set of data (dFORC diagrams as
a function of field rate and the waiting time in the reversal point). The
simple model presented here is in the same time of notable simplicity and
still is able to describe in a coherent manner the complexity of the
experimental data sets. We also mention the fact that we introduce a
methodology for characterizing rate dependent hysteretic processes with the
FORC diagram technique, which was used mostly for systems and experiments in
which the field rate influence was neglected even in cases in which this
effect can't be avoided and has a strong effect on the switching processes.
As many applications of the magnetic wires are dependent on the fast
switchings of the magnetization, we expect that this type of experiment to
be extensively used by researchers in order to observe systematically how
these processes are influenced by various physical parameters. This study
opens at least two new lines of research, one centered on the experimental
dynamic FORCs and the numerical treatment of experimental data and the
second on the model of the domain wall movement which can be dramatically
improved in order to include other elements that we have neglected in order
to maintain the clarity in a limited space of an article. We already have
identified elements from the FORC diagrams that can be correlated to
physical parameters of the wall movement (e.g., wall speed limit versus
field) and we are very confident that this method could become a valuable
tool to obtain important physical information about the dynamic
magnetization processes in a variety of magnetic wires and possibly be
extended to other magnetic systems.

\section{Acknowledgments}

Work was supported by Romanian CNCS-UEFISCDI Grants No.
PN-II-RU-TE-2012-3-0439 and No. PN-II-RU-TE-2012-3-0449.

\end{document}